# The Design and Implementation of a Quantum Information Science Undergraduate Program

Sarah Blanchette[1*], Danièle Normandin[1], Michel Pioro-Ladrière[2,4], Lyne St-Hilaire[1,2], Armand Soldera[1,3,4], Karl Thibault[4] and Dave Touchette[4,5]

[1]*Faculté des sciences, Université de Sherbrooke, Sherbrooke, QC, Canada*
[2]*Département de physique, Université de Sherbrooke, Sherbrooke, QC, Canada*
[3]*Département de chimie, Université de Sherbrooke, Sherbrooke, QC, Canada*
[4]*Institut quantique, Université de Sherbrooke, Sherbrooke, QC, Canada*
[5]*Département d'informatique, Université de Sherbrooke, Sherbrooke, QC, Canada*

Quantum information science is a burgeoning research field attracting vast public and private investment in the last decade. This quick rise has led to a talent gap, where there are more open positions than new graduates who can fill these roles. To meet this critical need, higher education has been challenged to react accordingly by assuring a flow of highly skilled individuals who must be trained quickly. We thus present how Université de Sherbrooke, in Quebec, Canada, responded by creating and launching an innovative undergraduate degree in quantum information science, aiming to address this gap by training quantum software developers in three and a half years. At the end of this program, they will be ready to join the quantum workforce. We detail the creative process leading to a coherent curriculum, as well as why the local ecosystem led to these choices. The guiding principles and lessons learned during the development of this interdepartmental and interfaculty degree are shared to inspire other quantum education institutions.

*Sarah.blanchette2@usherbrooke.ca

# 1. Introduction

Quantum information science (QIS) is the field which studies the processing, analysis, and transmission of quantum information used in quantum computers. Originally proposed in 1980s [1, 2, 3] the quantum computer has quickly evolved in the last decade from a theoretical concept to actual proof-of-concept devices. A fully operational quantum computer is widely believed to be achieved in the next decades; such a tool could impact our society in large ways, e.g. by decrypting all communications sent today using RSA encryption or by simulating molecules useful for pharmaceutical or industrial chemistry processes.

Due to its anticipated usefulness [4], giants in traditional computing and other industries have massively invested in quantum computing research and development (R&D), both in software and in hardware. Moreover, governments all around the world wish to retain expertise and access to this new technology, having thus put forward strategies surrounding the local and national development of quantum computing. This has led to the emergence of many quantum startups and the growth of existing tech companies, which require specific infrastructure and expertise.

This acceleration of R&D in quantum computing has faced a major bottleneck in the past few years: a lack of available highly qualified personnel [5]. This bottleneck partly stems from the fact that, historically, quantum information has been taught through PhD programs which output trained personnel at 28-34 years of age. Furthermore, these highly focused science PhDs may lack useful skills in an industry setting (e.g. communication, teamwork and/or product-oriented project management) while also lacking some knowledge or training in specific aspects of quantum information science.

Therefore, it is natural to ask whether it is possible to accelerate the training of highly qualified professionals in quantum information science. To address this, two aspects must be clarified. First, which subset of quantum information science should be prioritized for learning? The field is very diverse and multidisciplinary, encompassing domains such as hardware, software, physics, mathematics, and information science. Second, what is the expected time frame for completing this accelerated training? Both answers depend on already available training programs and job opportunities forecasted in each local ecosystem.

These issues can be distilled into a single pivotal question: "Can quantum software developers be trained within an undergraduate program of 3½ years?". At the Faculty of Science at Université de Sherbrooke, we initiated this reflection in 2019. This manuscript delves into the entire process and the associated challenges that led to the creation of this very new bachelor's degree in QIS. We hope that our approach provides valuable insights for other higher education institutions seeking to establish similar programs.

## 2. Program Development and Overview

To offer a clear understanding of the context surrounding the undergraduate degree in QIS, we begin with an overview of the quantum ecosystem in Sherbrooke. Next, we describe the input profile of students, highlighting the unique aspects of the Quebec education system. We then detail the interview process, a critical initial step in the program creation, and present the key findings which we call the three guiding principles Finally, we provide an outline of the undergraduate program itself.

### 2.1 Quantum in Sherbrooke

Starting in 2019, Université de Sherbrooke (UdeS) decided to address the workforce problem by proposing the creation of a new specialized bachelor's degree, which would end up being called the Quantum Information Science undergraduate program. It was launched in the fall of 2022. This degree is focused on quantum software development, as it was and still is believed the existing programs in physics and engineering provided sufficient training opportunities in quantum computing hardware. The physics' department at UdeS has cultivated expertise in quantum science for nearly 50 years, with a particular focus on quantum materials and quantum computing. This strong scientific legacy also paved the way for the establishment of Institut quantique (IQ) in 2016, an interdisciplinary research institute within UdeS. Today, IQ gathers 39 professors along with 254 graduate students of various departments and faculties, while also hosting the AlgoLab, a team of research professionals working on quantum software, amongst other initiatives. The creation of the undergraduate program in 2022 was strategically timed, aligning with the designation of Sherbrooke's Quantum Innovation Zone, Distriq, in the same year. Quebec's innovation zones are a provincial government initiative designed to help cities develop technology-driven ecosystems centered around their unique expertise. This approach fosters a vibrant environment where industry and academia intersect, attracting companies from around the world and accelerating start-up development. Today, Distriq operates a 50,000-square-foot collaborative environment dedicated to R&D, including offices, conference rooms, shared workspaces, open areas, access to technical expertise through its specialized and shared laboratory DevTeQ, while also hosting several startups. The new QIS undergraduate program is thus positioned within an already dynamic and rapidly expanding quantum ecosystem in Sherbrooke.

## 2.2 Creating the program

### 2.2.1 Background

To narrow down the question of whether highly qualified personnel in QIS can be trained in a shorter time frame, it is first important to examine existing training opportunities at Université de Sherbrooke. At that time, joining a PhD program was the primary way to join the quantum workforce in Sherbrooke. Specifically, the physics PhD program (either as an experimentalist or theorist) was the typical avenue, although it is now possible to pursue engineering, computer science, mathematics and chemistry PhDs in Sherbrooke related to QIS. The major drawback of relying on such highly specific training programs is their length: as mentioned, a PhD student typically dedicates nearly a decade to higher education. Moreover, such highly specific expertise alone does not align with the diverse needs of the emerging quantum industry as we will see in Section 2.2.2. A 2022 study by the Quantum Economic Development Consortium based on an earlier survey also showed that most jobs in the field do not require a PhD [6]. An alternative pathway to join the quantum workforce has been through specialized undergraduate programs in fields related to QIS. The most prominent one at the time was the program in electrical engineering, which was suited to a quantum hardware-oriented future. However, no undergraduate program at UdeS clearly prepared students for software-oriented roles in quantum technology. The program we present thus aims to address that gap, training quantum programmers capable of mapping practical problems to algorithms that can run on quantum computers.

With the student output profile briefly established, we now turn to the unique aspects of the Québec education system, which sets it apart from other North American systems. Understanding this context will help clarify the background of incoming students and thus the rationale behind the program structure. In Quebec, students go through two years of *CEGEP* after five high school years before pursuing undergraduate studies in university. Table 1 illustrates the full educational timeline from primary school to undergraduate studies. Note that the typical duration of science undergraduate programs is 3 years, often extended to 3 ½ years for the coop regime (Section 4.4), and 4 ½ years for coop engineering undergraduate programs.

*Table 1: Québec's education system timeline*

| 6 yrs | 5 yrs | 2 yrs | 3 yrs |
|---|---|---|---|
| Primary | High School | CEGEP | Undergrad |

*CEGEPs* offer two major kinds of programs: those that prepare students for university and those that lead directly to the job market, such as technical training programs, similar to trade school. Among the most popular pre-university programs are Sciences, Computer Science and Mathematics (SCSM), as well as Natural Sciences. The latter is split into two tracks: Health Science and Pure Sciences. These programs include three physics courses, namely, Classical Mechanics, Electricity and Magnetism, and Waves & Modern Physics, as well as three math courses: Calculus 1&2 and Linear Algebra. This information will be referenced again in [Section 3](#). For now, it is sufficient to mention that students entering the QIS program have an understanding of classical mechanics, electromagnetism and optics, derivatives and integrals and linear algebra, with the SCSM students also possessing basic coding skills.

### 2.2.2 Interviews

With the overall output and input profiles of students in this new QIS program established, the next stage is to detail skills and knowledge needed to work in this field, whether in academia or industry. However, for emerging fields like quantum sciences, determining key skills is a fundamental and far-reaching step in creating comprehensive and adequate training programs [7]. To this end, close to 50 interviews were conducted back in 2020, 30 with representatives from large companies and startups, both local and international, and 20 with members of the student community.

To better understand the needs of the quantum industry, we focused on the following four questions:

- What are the key skills to work in the quantum field?
- What are the skills currently lacking or difficult to find in the quantum field?
- What would be the benefits of hiring someone with an undergraduate degree versus a graduate student?
- What are the pros and cons of hiring interns?

Importantly, the interviewees were unaware of the intention to focus on quantum software. These questions were asked as generally as possible to gather unbiased insights. Here is a summary of the key results:

- Required competencies to work in the quantum industry include programming skills (good code structure), project management, teamwork abilities and a solid foundation in quantum related fields, highlighting the need for multidisciplinarity.
- The most challenging skills to find are robust coding skills, a multidisciplinary approach, problem-driven mindset, project management and teamwork abilities.

- The current lack of talent in the field justifies hiring undergrad holders, especially for larger companies, while start-ups tend to prefer candidates with a very specific expertise.
- Internships are seen as valuable but are less beneficial if limited to only four months. Longer internships should be privileged to allow for deeper learning and contribution.

Equally important were the interviews conducted with the student community to gauge interest. We targeted potential incoming students, as well as current students and recent graduates in quantum science related programs in Quebec and France. These interviews identified characteristics of an attractive program, including multiple potential career paths, a strong emphasis on practical training and cross-cutting professional skills. The detailed questions and responses of these interviews can be found in the [Appendix](Appendix).

The combined findings from industry and student interviews shaped three guiding principles for the undergraduate program in QIS:

1. Train a new generation of interdisciplinary scientists ([Section 3](Section 3));

2. Implement an innovative and interdisciplinary pedagogical approach ([Section 4](Section 4));

3. Provide a professional, student-centered, training program ([Section 4](Section 4)).

These principles have guided the selection and arrangement of pedagogical activities, the organization of course content, the inclusion of mandatory internships, and the modalities of learning, teaching, supervision, and evaluation that will form the backbone of this undergraduate degree.

### 2.2.3 Team set-up

In addition to insights gained from the many interviews, UdeS set up a development team and an expert committee. The former was composed of the Operations and Talent development officer of Institut quantique, the coordinator of Development & Partnerships of the Faculty of Science and the Project coordinator of the *Accompagnateur entrepreneurial Desjardins.* These three resources worked part time to this end over the course of two years. The expert committee, on its end, was composed of professors from the departments of Physics, Mathematics, Computer Science and Biology of the Faculty of Science, professors from the Faculty of Engineering and the Faculty of Management, as well as the executive director of Institut quantique. This group was consulted periodically throughout the creation period, to determine the required pedagogical activities and the program's direction.

## 2.3 Program overview

The curriculum development began with a focus on identifying the fundamental scientific knowledge deemed essential. The training needed to be sufficiently robust in scientific fundamentals to ensure that graduates could continue to track the developments in the field after completing their degree. The scientific foundations in quantum computing, mathematics, computer science and physics, make up the majority of courses students will undertake.

After establishing these bases, the team identified existing activities best suited to cover these concepts, and then developed new activities considered necessary to complement the training. The bridge between these scientific disciplines is the integrative project courses, detailed in Section 4.4. The program also benefits from courses in the faculties of Engineering and Management.

Industry needs and student aspirations are reflected throughout the program in various ways, such as through the three mandatory internships and professional development courses, which benefit both students and the future employers.

Figure 1 outlines how each program objective is supported by corresponding pedagogical activities. They have been designed to prepare graduates for three main career paths: 1) careers in industry, 2) entrepreneurial careers, and 3) advanced studies leading to careers in academia or research labs.

**Course legend**:

| | |
|---|---|
| PHQ | Offered by the Physics department |
| MAT & STT | Offered by the Mathematics department |
| IFT | Offered by the Computer Science department |
| GEI | Offered by the Faculty of Engineering |
| BSQ & SCI | Offered by the Faculty of Science |

|  | Year 1 | | | Year 2 | | | Year 3 | | | Year 4 |
|---|---|---|---|---|---|---|---|---|---|---|
|  | S1 | S2 | W0 | S3 | S4 | W1 | S5 | W2 | W3 | S6 |
| **Understanding the required mathematics for the analysis of quantum phenomena** | MAT120 (3cr.) Discrete Mathematics<br>MAT193 (3 cr.) Linear Algebra | MAT298 (3 cr.) Vector Calculus | | STT290 (3 cr.) Probabilities | | | MAT189 (3 cr.) Real Analysis | | | Four elective courses, chosen according to the career plan, for a total of 12 credits subject to approval by the program committee.<br><br>Note that two more elective courses (not shown here) are also in the course grid, one in S4 and another in S5. |
| **Mastering classical programming and algorithms** | SCI102 (2cr.) Scientific Tools<br>IFT159 (3 cr.) Computer Programming and Analysis | IFT232 (3 cr.) Object-Oriented Design<br>IFT339 (3 cr.) Data Structures | | IFT436 (3 cr.) Algorithms and Data Structures | IFT503 (3 cr.) Theory of Computation | | | | | |
| **Knowing how to model quantum systems** | | PHQ230 (3 cr.) Quantum Mechanics | | PHQ401 (3 cr.) Physics of Quantum Systems | PHQ404 (3 cr.) Numerical Methods and Simulations | | PHQ476 (3 cr.) Physics of Information<br>PHQ598 (3 cr.) Architectures of Quantum Computers | | | |
| **Mastering the basics of quantum computing** | BSQ112 (3 cr.) Introduction to Quantum Computing | | | | PHQ533 (3 cr.) Quantum Information and Quantum Computation | | | | | |
| **Develop and integrate professional competencies needed to work in the field of quantum science and technology** | | BSQ101 (3 cr.) Integrative Projects in Quantum Programming | | GEI299 (2 cr.) Basic Design and Management for Technology Projects<br>BSQ201 (3 cr.) Integrative Projects in Quantum Solutions | BSQ301 (3 cr.) Integrative Projects in Science Popularization | | BSQ401 (3 cr.) Entrepreneurial Projects in Quantum Science and Technology | | | BSQ501 (3 cr.) Specialty Project in Quantum Science |
| **Defining one's own expertise, take one's place among disciplinary specialists and adapting to evolution in the field** | BSQ111 (1 cr.) Professional Development: Professional Plan in Quantum Sciences | | | BSQ222 (1 cr.) Professional Development : Competencies, Knowledge of Internships Settings and Ethical Considerations | | | BSQ510 (2 cr.) Opportunities in Quantum Science and Technology | BSQ333 (1 cr.) Professional Development : Career Plan in Quantum Science | | |

*Figure 1: Teaching objectives/Teaching activities and program chronology. The program curriculum is divided into two main sections. The first (in blue), regroups the science courses, whereas the second (in pink) regroups courses focused on the development of personal and professional skills The semesters tagged W1, W2 and W3 are the three mandatory work terms.*

# 3. A new generation of interdisciplinary scientists

The primary guiding principle put forward in [Section 2](Section 2) is the training of a new generation of interdisciplinary scientists. This section details how the three scientific foundational disciplines are included into the program and demonstrates how they interconnect to help students develop a comprehensive understanding of the quantum field within each discipline.

## 3.1 Mathematics

Mathematics is fundamental to quantum information theory as it provides the essential language and tools to describe quantum systems and algorithms. Key concepts from linear algebra, including vector spaces and matrices, are essential for understanding quantum states and operations, while complex numbers and probability theory are crucial to addressing the inherently probabilistic nature of quantum mechanics. Advanced mathematical techniques are also required to develop algorithms and protocols for quantum computing and communication.

As shown in [Figure 1](Figure 1), the program includes four math courses within the first three semesters, similar to other science programs, where foundational courses are introduced early. The courses in linear algebra, probability, discrete math and vector calculus are taken from the math curriculum, while the real analysis class in the fifth semester is tailored specifically to meet the needs of our program.

## 3.2 Physics

Physics is a cornerstone in the program because it provides the fundamental principles governing quantum systems. Quantum mechanics describes the behavior of particles at the smallest scales, which is essential for understanding the functioning of quantum bits (qubits). The program's final course grid includes six carefully selected physics courses, while fundamental courses such as Classical Mechanics, Waves and Modern Physics, and Electromagnetism (EM) are intentionally omitted, as students have already encountered them at an introductory level in CEGEP ([Section 2.2.1](Section 2.2.1)). A specific selection of electives is proposed to students planning to pursue graduate studies in physics, which require a university-level EM course.

Among the physics courses in the curriculum, some are newly developed, others are shared with the physics program, and a few have been adapted to QIS students. For example, the first Quantum Mechanics (QM) course was redesigned to align with current recommendations and is structured around spin systems [8]. Fundamental topics, such as quantum tunneling and infinite potential wells, are introduced during the two last weeks of the semester, contrary to the usual QM course where these topics are introduced first. The whole course is taught from a quantum computing perspective: two-level systems are presented as qubits, and time evolution is associated with gates applied on a quantum circuit. The second Quantum Mechanics course, *Physics of Quantum Systems*, introduces students to topics like the harmonic oscillator, coherent, cat and GKP states. Additionally, students explore the Jaynes-Cummings Hamiltonian, variational methods, identical particles, the Jordan-Wigner mapping, and time-independent perturbation theory. This course is different than the traditional *Quantum Mechanics II* course taught in the physics curriculum, which also covers topics such as the hydrogen atom and the angular momentum operator, offering a tailored approach more relevant to quantum information science.

### 3.3 Computer Science

Computer science plays a critical role in the development of quantum algorithms and computational models, focusing on designing and analyzing algorithms that leverage the unique capabilities of quantum computers. Key concepts from theoretical computer science, including algorithm design, computational complexity and information theory, are crucial in understanding the potential and limitations of quantum computing. Computer science also contributes to the development of quantum programming languages and software tools required for implementing and testing quantum algorithms. Moreover, industry feedback has highlighted strong coding skills as a top priority, underscoring the need for practical experience. In response, the program offers six computer science courses to equip students with these foundational skills.

One of these courses, introduced in the first semester, teaches Python, a prominent language for quantum computing, as well as different collaborative tools that students will use extensively in many integrative project courses. Simultaneously, students begin a sequence of four core courses from the traditional computer science curriculum, designed to build a solid foundation in coding and algorithmic design. Additionally, a course on the theory of computation, typically an elective in the traditional computer science curriculum, has been deemed mandatory in our curriculum to provide essential context on the capabilities of quantum computers relative to traditional computation models.

### 3.4 Linking everything together

The term interdisciplinarity suggests a common space that unifies different areas of knowledge, fostering cohesion across fields [9]. Given the inherently interdisciplinary nature of quantum sciences, the program must not only provide the scientific foundations outlined above, but also integrate them to form a coherent set of competencies for students. This integration begins in the first semester with the course *Introduction to Quantum Computing*, where students are exposed to the basics of quantum computing. In this course, they apply relevant mathematical tools such as linear algebra and complex numbers to learn about qubits and quantum mechanical phenomena like superposition. The course also introduces students to quantum programming, providing an essential foundation for future studies in the field.

Some courses in this program are exclusive to QIS students, but for most of the classes, they join existing cohorts, fostering a beneficial diversity within the classroom. This is advantageous for both the new program and existing ones: a classroom with students from various academic backgrounds promotes debates, introduces new perspectives, raises stimulating questions, nurtures curiosity, and enriches team projects. Additionally, by interacting with students from traditional disciplinary programs, quantum science students gain a deeper appreciation of the unique and fundamentally interdisciplinary nature of their own field.

The integrative project courses, which will be discussed in the next section, offer an excellent opportunity for students to apply notions seen in the disciplinary courses, reinforcing their mastery of the basics of quantum information.

Graduates will be scientifically well-equipped to effectively interact at the interfaces of disciplines related to quantum sciences, regardless of the professional context they pursue. Although quantum sciences are an emerging and rapidly evolving field, the program ensures a solid scientific foundation, equipping students to understand and integrate technological advancements as they arise in the coming years.

# 4. An innovative pedagogical approach leading to a professional training

The second guiding principle outlined in [Section 2](#) is the use of a pedagogical approach that promotes the breakdown of traditional disciplinary silos, enabling the appropriation and integration of scientific knowledge so that students can develop their own interdisciplinary expertise. Indeed, a major challenge for a program spanning various disciplines is to ensure continuity in learning and facilitate connections across these disciplinary courses. To meet this challenge, we have incorporated activities that allow students to apply their acquired knowledge and skills to solve real-world problems, providing opportunities to observe and experience the inherently interdisciplinary nature of quantum sciences. We refer to these activities as integrative projects.

The third guiding principle is to offer a professionally oriented degree that enables graduates to pursue employment in their field upon program completion. Although integrative projects serve this purpose, the program also includes specific activities such as courses in personal and professional development, as well as a dedicated course designed to help students explore different career opportunities post-graduation. These courses will be detailed in this section. They are especially relevant in a specialized discipline like quantum computing, where many students begin the program with limited knowledge of the specific competencies, such as the ability to communicate effectively with people from different technical backgrounds, which are required for careers in the field.

## 4.1 Student-centered pedagogy

The phrase "I teach less, they learn more" [10] reflects a pedagogical approach that prioritizes active student learning. Rather than relying on traditional teaching where the instructor primarily transmits knowledge, this method emphasizes the active participation of students in the learning process.

By teaching less, the instructor fosters an environment where students are encouraged to explore, ask questions, and solve problems on their own. This includes activities such as group discussions, collaborative projects, and independent research, all of which contribute to developing essential transversal skills like analytical thinking, creativity, and autonomy. In this model, students become the main actors of their own learning. Ultimately, by reducing the time spent on direct instruction and increasing opportunities for active learning, students can achieve a deeper and more lasting understanding of the subjects they study [11]. Our pedagogical approach aims to innovate by creating space for students to actively construct knowledge with the support of peers and instructors. This approach reflects a shift towards student-centered learning, moving away from traditional lecture-based methods. By adopting these pedagogical methods, educators seek to create a learning environment that involves students as active participants, helping them to take ownership of their educational journey.

### 4.2 Personal and professional development

Another innovative aspect of the program is the emphasis on personal and professional development. To foster the development of cross-cutting skills, three one-credit courses on personal and professional development have been created. The first course focuses on social skills (communication, kindness, organization, teamwork), as well as the place of quantum science and scientists in society. The second course covers knowledge of internship environments and the expected skills from an employer in quantum computing (autonomy, ethical responsibility), whereas the third course encourages students to think about their future choices, develop their creativity, and take their place in the quantum ecosystem. The reflection work is spread out over the three years, during which individuals build a skills portfolio which includes analysis of their competencies. The portfolio is a great tool for learning about oneself, understanding one's strengths and challenges and it is also a great way to prepare for the job market. These courses distinguish the program from regular science programs, and are an important attribute in making the program professionalizing.

Another course which isn't tagged as a development course but fits into the broader context of professionalization is *Opportunities in Quantum Sciences and Technology*. It is a two credit course, strategically placed before the 8-month internship (W2 and W3 in [Figure 1](#)), whose goal is to offer students tools to seize opportunities (e.g., how to stand out, networking tips, etc.) and to present the possible career paths upon graduation.

### 4.3 Integrative project courses

As shown in [Figure 1](#), the program has five integrative project (IP) courses. Beyond their aims of application and integration of disciplinary knowledge, these projects foster the development of professional skills related to autonomy, teamwork, communication, and project management. Additionally, they incorporate a reflective dimension, prompting the group to question their preconceptions, pose hypotheses, and validate them within the framework of the ongoing project. This type of approach empowers students to constructively question their own understanding and that of their peers, which proves to be a crucial skill in such an innovative field as quantum sciences. Students work in teams on projects spanning from three weeks to a full semester, gaining experience in collaborative coding, writing reports, presenting their work in front of peers, and in developing their leadership and business-oriented mindsets. The idea for these courses is also for students to interact with different groups outside of university walls, especially starting in their 2nd year. The courses are complementary in nature and are adapted to students' progress through the curriculum.

The first of these courses, *IP in Quantum Programming*, includes smaller projects to help students acclimate to this course format. The shorter project duration also enables the inclusion of more concepts, which are discussed in the [Appendix](Appendix), along with all other courses presented here. In the following semester, the *IP in Quantum Solutions* tasks students to focus on industrial use-cases, considering quantum and classical solutions. Through two 6-week projects, students gain deeper understanding of a chosen topic and practice project management methods. This course includes concepts learned in the physics, math and computer science courses, and incorporates teachings from the engineering course *Basic Design and Management for Technology Projects*, also offered in the 3$^{rd}$ semester. This lecture-based course leverages the Faculty of Engineering's expertise in technical project management, adding substantial value to the curriculum. In the 4$^{th}$ semester, the *IP in Science Popularization* course, allows students to explore different media forms to communicate quantum technology concepts to different target audiences. Shifting focus, the fourth integrative project, *Entrepreneurial Project in Quantum Science and Technology*, emphasizes entrepreneurship, building on students' maturity and project management experience. Finally, students can personalize their last IP course to align with their intended career path: research, academia, industry, entrepreneurship, etc. This customization supports the program's last semester as a year of specialization, a topic further discussed in the next subsection.

### 4.4 Internships

A key feature of the QIS program is its cooperative (co-op) structure, which positions it as a professionalizing training program. Under this model, students alternate between regular class semesters and paid work placements, as can be seen in [Figure 1,](Figure 1) providing them with both theoretical knowledge and hands-on experience. These internships, highly desired by industry interviewees, facilitate professional integration while allowing students to apply their newly acquired knowledge and skills in a real-world context. Internships account for at least 30% of the program, meaning that QIS graduates already have a year of relevant experience in the field by graduation. The first mandatory internship takes place after the 4th semester, or two years of the program. Students are encouraged to find an internship in classical software development during this internship, as for most of them this is their first *professional* job. The second and third internship are combined and come after the 5th semester. While less common at the university, this extended internship meets industry preferences for longer placements, allowing students to engage deeply with their project and become familiar with the company's technology. This extended term also international internships, allowing students to potentially complete their final year abroad and taking the last courses in a partner university.

In their final year, students can tailor their studies to specialize. Indeed, this semester allows each student to tailor their training with 12 credits of elective courses adapted to their specific career profile. This flexibility is especially relevant as they return from significant industry experience. If a company hires an intern and plans to bring them on full-time after graduation, the final semester serves as an ideal opportunity for students to gain additional knowledge that will benefit them as an employee.

## 5. Ensuring the program's success

Throughout the development and launch of the program, a series of best practices were established. The first subsection highlights approaches used in the program's creation process, while the second and third subsections put forward current practices to ensure the program remains relevant and valuable to all stakeholders.

### 5.1 Overcoming hurdles

The creation of a new QIS undergraduate degree required strong institutional support, led by the Vice-Dean for Development and Partnerships of the Faculty of Science. This support was essential to stimulate discussions throughout different departments and establish a unified vision for developing the program content. Recognizing and leveraging the existing expertise within the three departments of the Faculty involved in this project was the next step, which presented several key challenges. Firstly, interdisciplinary collaboration among researchers often centers around a shared theme. In this scenario, it was crucial to unite diverse areas of expertise to build a cohesive and innovative educational program. Second, differences emerged regarding the approach to certain scientific concepts, with each specialty having its own perspectives and methodologies. This required aligning distinct scientific perspectives and methodologies unique to each specialty. Constructive, open dialogue in frequent meetings fostered synergy between fields, helping bridge conceptual differences. Strong leadership from the Dean's office, coupled with a strict two-year timeline, helped resolve conflicts swiftly. In sum, this collaborative approach enabled the design of an educational program that integrates the strengths and expertise of each department, establishing a solid foundation for QIS training.

### 5.2 Feedback from collaborators

In the first years of a new program, it is essential to get feedback from every university stakeholder: students, professors and lecturers as well as the Internship and Professional Development Department (IPDD). One mechanism for doing so is through the QIS program committee, which meets at least once every semester. Its members include the QIS program leadership and coordination team, representatives from each department involved, a member from the IPDD, the Faculty of Science's pedagogical advisor, and one student representative per cohort. These sessions allow each contributor to stay updated on overall program developments and exchange insights. Monthly meetings with the teaching staff are also valuable for fostering course cohesion, especially for initial offerings of IP classes, where aligning with disciplinary content is a central objective. Through these meetings, many changes have been made to the original program. Following student feedback on the relevance of some courses, a few have been removed and replaced, while others have been moved from one semester to another.

### 5.3 Staying in touch with industry

As the field is rapidly evolving and the program has a professional focus, staying engaged with industry developments is essential. The program has established partnerships with several quantum computing companies, and a member of the direction team is partially dedicated to managing these connections. These partnerships allow faculty to collaborate with industry via projects, incorporate specific software development kits or libraries into courses, and in some cases, offer students exclusive access to quantum hardware. Such contacts also provide complementary expertise into the classroom, and local quantum companies frequently contribute through guest lectures on both scientific topics and industry insights. Maintaining strong ties with industry is a core aspect of the program, ensuring its relevance and alignment with the evolving demand for skilled talent, while also connecting students to the ecosystem.

## 6. Conclusion

In this article, we have outlined the creative process behind the development of the Quantum Information Science undergraduate program at Université de Sherbrooke, along with its implementation. We explained the rationale for establishing this specific program, driven by the needs of the local ecosystem. Since its launch in 2022, the program has grown to have three active cohorts with over 50 enrolled students. Student feedback has been overwhelmingly positive, with a particular appreciation for the integrative project courses. These courses have significantly enhanced students teamwork and overall communication skills, demonstrating the effectiveness of the many mechanisms put in place to foster these competencies. As of fall 2024, the first cohort has ended their first mandatory internships, with students having worked at governmental agencies, academic institutions, research institutes, local startups, and multinational corporations. We hope our success story can inspire other higher education institutions to develop similar programs, helping to close the talent gap in quantum computing.

## 7. Acknowledgements

The authors would like to thank Émilie Desrosiers, Maxime Dion, Stéphanie Perron and the Expert Committee consulted during the creation process for their time and valuable input: Pr. Yves Bérubé-Lauzière, Pr. Jean Bibeau, Pr. Claude Bourbonnais, Pr. Stefanos Kourtis, Pre. Isabelle Laforest-Lapointe, Pr. Denis Morris, Christian Sarra-Bournet and Pre. Vasilisa Shramchenko. We are grateful for the institutional support by the rectorate, specifically Pre. Christine Hudon, as well as by the deanery of science, specifically to Pre. Carole Beaulieu and Pre. Nancy Dumais.

Thank you to Institut quantique for their support, the team at the AED, as well to all the implicated departments and faculties that make this program happen. We appreciate the invaluable support of Vicky Richard and Catherine Vallières on the project booklet for the creation of a bachelor's degree in quantum sciences, submitted for evaluation by the Bureau de Coopération Interuniversitaire (BCI). We woud also like to thank the BCI and the Québec's Ministère de l'Enseignement Supérieur to have accepted and believed in the project. Finally, the authors would like to acknowledge the interviewees for their helpful participation as well as the interviewers, Élie Génois, Ioanna Kriekouki, Maxime Lapointe-Major, Camille Le Calonnec, Charles Paradis and Patrick Bourgeois-Hope.## Bibliography

# Appendix 1

## Interview results

<u>With the student community</u>
The interviewed people were current students or recently graduated students of many institutions: CEGEPs throughout Québec, Université de Sherbrooke's Faculty of Science and Faculty of Engineering, Université Laval and Université de Strasbourg.

The questions asked were the following:

1. What is the pre-university community's vision of quantum?
2. What reasons motivate the student community to choose a Bachelors' degree in Science?
3. What are missing skills employees in quantum would have liked to gain during their studies?
4. What is the relevance of internships during one's studies?

The principal conclusions go as follow:

**Perception of quantum sciences:**
- Quantum science and technology appears inaccessible by pre-university students working in the field, and most of them know close to nothing about the field.
- The way most pre-university physics classes are taught don't seem to generate much interest the field of quantum.

**Bachelors' degree in Science**
- People choosing a BSc. often do so to insure a many options of career paths. Many students expressed that a BSc. shouldn't lead to a single output, but should allow for many opportunities.
- Engineering students said they have a strong interest in science, but chose engineering because it appears more concrete.

**Skills lacking when joining the workforce**
- Science programs generally leave little room for the development of practical skills. Participants mentioned classes should have less theoretical homework and more project-based homework.
- Sciences programs are often designed to cover too broad a spectrum of knowledge. As a result, many told us they had gaps in their basic knowledge, and were looking for a program that provided a solid foundation.

- Transverse skills (communication, project management, etc...) are often cited as being essential in every field of work, but are lacking in science training.

**Relevance of internships during studies :**
- Internships are clearly seen as an essential part of the training program.
- Internships could be offered for teams of two or three people, which could encourage teamwork and increase the impact of internships for industries (two or three interns working in a team offer much more value than two internships with no interaction).

# Appendix 2

## Integrative projects

**BSQ101 - IPs in Quantum Programming:** This course is students' first opportunity to apply notions learned in their math, physics and computer sciences courses, finally exploring quantum computing's multidisciplinary aspects. Four three-week projects are presented throughout the semester. The challenges are always presented and discussed in class, and then students work in teams to implement a solution based on a quantum algorithm. For example, a Boolean satisfiability problem is presented and students must use Grover's algorithm to solve it. The complete course cursus can be found in French [here](#).

**BSQ201 - IPs in Quantum Solutions**: The idea behind this course is to give students industrial use-cases and have them provide both classical and quantum solutions. It is also an opportunity for them to be in contact with the participating stakeholders. In the first iterations of the course, students had a first project on quantum machine learning and the second on a quantum adiabatic algorithm using neutral-atom hardware. The complete course cursus can be found in French [here](#).

**BSQ301 - IPs in Science popularization**: The four projects are divided as follows: written communication, communication via comics, participative communication (quiz, game, etc.) and oral communication. The last two projects were combined and presented in a local CEGEP to science students. The complete course cursus can be found in French [here](#).

**BSQ401 - Entrepreneurial Project in Quantum Science and Technology**: The goal is for students to develop an entrepreneurial mindset, attitude and language by taking a quantum technology project idea and turning it into a business model. This course is given in collaboration with the Faculty of Management and the ac

**BSQ501 - Specialty Project in Quantum Sciences**: This course is quite different from the others, as each student will have their own personalized project. Students will be coached according to their desired career profile.